\documentstyle[12pt,a41,epsfig,floatfig,psfig,rotating]{article}

\newcounter{my}

\newcommand{\la}[1]{\label{#1}}
\newcommand{\re}[1]{\ (\ref{#1})}
\newcommand{\nn}{\nonumber}
\newcommand{\ed}{\end{document}}
\newcommand{\be}{\begin{equation}}
\newcommand{\ee}{\end{equation}}

\newcommand{\ba}{\begin{eqnarray}}
\newcommand{\ea}{\end{eqnarray}}
\newcommand{\baz}{\begin{eqnarray*}}
\newcommand{\eaz}{\end{eqnarray*}}
\newcommand{\bb}{\begin{thebibliography}}
\newcommand{\eb}{\end{thebibliography}}
\newcommand{\ct}[1]{${\cite{#1}}$}

\newcommand{\bi}[1]{\bibitem{#1}}

\begin{document}

\initfloatingfigs
\sloppy
\thispagestyle{empty}

\begin{center}
{\LARGE \bf  Instantons in the
 Large  {\boldmath $Q^2$, $x$}  Region at HERA}

\vspace{5mm}
 Nikolai Kochelev$^{1}$ and Albert De Roeck $^2$\\
\vspace{5mm}
{\small\it
(1)JINR, Dubna, Moscow region, 141980, Russia\\
(2) DESY,  Hamburg, Germany, and CERN, Geneva, Switzerland}\\

\vspace*{2cm}
\end{center}
\begin{abstract}
\noindent
A new mechanism for an increased event production
in the large  $Q^2$, $x$ region at HERA is suggested.
This mechanism results from  a  new type of instanton-induced
quark-quark interaction, which is  related  to  
non-zero  quark modes  in the instanton
field. We estimate the contribution of this interaction to the 
valence quark structure function $F_2^v$ using the  
gas instanton model approximation for
the QCD vacuum. 
It is shown that this interaction can give a large contribution
to $F_2^v$, especially at large values of $Q^2$. The strong 
dependence of this contribution on the quark masses and $Q^2$
is discussed, and it is concluded that  a large charm quark contribution 
 to events at high $Q^2$ can result from
  charm pair  creation in the instanton field. We also obtain  a
sizeable contribution to the valence quark longitudinal  
structure function $F_L^v$
at high $Q^2$.
Experimental observables  which can  check this instanton mechanism in the
high $Q^2$ region at HERA
are  suggested.

\end{abstract}
\newpage
\section{Introduction}
\vspace{1mm}
\noindent

Recently\ct{HERA} the H1 and ZEUS collaborations at the 
electron proton collider HERA announced an 
intriguing excess of events above Standard Model expectation in the
deep inelastic scattering (DIS)
region of large $Q^2$ $(> 10000 GeV^2)$ and large $x$.
When confirmed by future data\footnote{with the additional data of the 1997
run the reported excess is somewhat 
reduced to a two sigma significance effect
\ct{bruell}}, this excess is likely to be the most exciting
result in particle physics of the last years.
 Many interesting explanations of the this excess, mostly exploring
physics scenarios
beyond Standard Model, have been suggested (see for example \ct{leptq}).

Within the Standard Model the largest uncertainty on the cross section
prediction results  from the assumed  parton densities in the region
of large $x$ and large $Q^2$. These are based on two main ingredients. 
Firstly, the parton densities are parametrized with a relatively
small amount of parameters at some low $Q^2$ scale, using  a fit based
on pQCD evolution equations of
 experimental data on lepton--parton and parton--parton cross sections at 
 smaller values of
$Q^2$. Secondly, to obtain a prediction
for the parton densities at $Q^2>10000GeV^2$, leading
twist (DGLAP) evolution equations
\ct{AP} are used.

A weakness of the first point was pointed out in \ct{tung} and \ct{charm}.
In~\ct{tung} it was shown  that, not unexpectedly,  the perturbative
QCD evolution generates a feed down mechanism, i.e. effects at very large
$x$ and small $Q^2$ values propagate to smaller $x$ at larger $Q^2$ values. 
Hence e.g. an 
unexpected hard component to the  partonic distributions at $x>0.75$ and
$Q^2\sim 20 GeV^2$ has a visible effect in the region where HERA 
reports an excess.
In~\ct{charm} the possible importance of an
intrinsic charm component in the proton wave function \ct{brodsky} to
explain the excess was pointed out.

However, from our point of view,  an important aspect of 
QCD has not been taken into account so far, namely the   role
of the complicated structure of the QCD vacuum related
 to the presence  of
strong nonperturbative fluctuations of gluon fields, 
called instantons \ct{Pol}.
 In \ct{kochhq} a discussion was  presented on
a possible instanton contribution  to the high $Q^2$ region at HERA.
It was found that instanton induced
interactions can enhance  the expected event rate at high
$Q^2$, resulting from 
 multiple emission of gluons at
the instanton vertex \ct{bal}.

Instantons describe the tunnelling between different
gauge rotated classical vacua in  QCD 
field theory, and reflect the nonabelian character of the theory
of strong interactions
\footnote{ For an introduction to the instanton physics 
see e.g. \ct{nsvz}.}.
In QCD  instantons play an important role in
chiral symmetry breaking and  in the origin of the masses of constituent
quarks and hadrons. In recent reviews \ct{shur}, \ct{diak} it was shown that
instantons can reproduce  fundamental quantities of the QCD vacuum,
such as the values of the 
different quark and gluon condensates, and can also give 
a good description of the hadron spectrum.

Particularly  interesting  are  the interactions which are 
connected with the so called   quark--quark
t'Hooft  \ct{Hooft} interactions, which are 
induced by
instantons. Four important  features of these interactions should be stressed.

One feature is the flavor dependence of the interaction, which leads 
 to  large contributions in flavour
singlet channels. For example,  the quark--antiquark
repulsion in the t'Hooft interaction for $SU(3)_f$ singlet $q\bar q$
states leads to a large enhancement
of the $\eta^\prime$ mass
and offers a  solution to the  $U_A(1)$ problem
\ct{Hooft}.
 On the other hand,  the t'Hooft interaction for
quark-quark attraction in a flavor
singlet state leads to the formation
of an isoscalar diquark state inside the nucleon  \ct{qsr}.
This has  direct relevance to DIS: it offers a natural
way to explain the experimental data on the ratio of proton to neutron
structure functions $F_2^n(x,Q^2)/F_2^p(x,Q^2)$ at large $x>0.1$, when
 assuming the presence of 
 a rather large isoscalar diquark component in the nucleon,
which does not disappear for increasing  $Q^2$ \ct{diquark}.

A second feature of the t'Hooft interaction is it's strong
helicity dependence: contrary to the perturbative quark--gluon
vertex, which conserves the quark helicity, instanton--induced quark--quark
 vertices exhibit a spin flip
and consequently lead to quark helicity violation effects. In 
\ct{kochprd,Forte}) it was shown that 
this contribution may explain the observed large
violation of the Ellis--Jaffe sum rule for the first moment of
spin--dependent structure function $g_1(x,Q^2)$ \ct{EJ}.

A further property is the  strong dependence 
of the t'Hooft interaction on the 
 mass of the quarks involved.
For the so called zero quark  modes (zero energy states of quarks in the 
instanton field) the contribution behaves as 
 $1/{m_q^*}^2exp(-2m_q^*\rho_c)$, where $m_q^*$ is the effective quark mass
in the instanton vacuum
and $\rho_c\approx 1.6GeV^{-1}$ is the average instanton size in 
vacuum\ct{shur}.
This leads to a large suppression of the 
 contribution of zero quark modes to the quark distribution function
for heavy flavours. A simple estimation in the framework of
the instanton liquid model \ct{shur}
shows that sea charm quarks are suppressed in  $F_2$ by 
a factor $10^{-4}$ compared to light sea quarks, and can therefore
be neglected.

A final feature is related to the strong dependence of the t'Hooft 
interaction on the  virtuality of the interacting quark.
The t'Hooft interaction results from the existence of
zero quark modes in instanton fields, hence they are strongly localized  
 in the field, which leads to an exponential dependence
of their Green function  on the quark virtuality.

One can calculate the instanton contribution to the sea quark distribution
function by considering the interaction of a
 quark  from  a virtual photon with a valence quark
in the nucleon through an instanton fluctuation (see Fig.~1). In this case all
properties
 of the instanton contribution are determined by the dependence of the
instanton--induced  quark--quark cross section on the virtuality of the
incoming sea quark.
Using  the model of  Landshoff~\ct{land}
one can show that a fast reduction of the strength of this interaction
with increasing  sea quark virtuality results into
 a leading twist  contribution in the sea quark distribution
function \ct{kochprd} (see also \ct{ring}).

The exponential dependence of the 
quark-instanton vertex for zero quark modes 
on the incoming  quark virtuality is expected
to lead to
a rather soft instanton--induced 
quark distribution as function of  Bjorken $x$.
This makes it difficult to explain a substantial  excess  
at large values of $x$ in the HERA region
   from quark zero modes 
in the instanton field (see \ct{kochhq}) alone.

The main goal of  this study  is to consider  the contribution of
so called
{\it non-zero  quark modes} in the instanton field 
(i.e. non-zero energy states of quarks in the 
instanton field) 
to  the $F_2(x,Q^2)$ and  $F_L(x,Q^2)$ structure functions, with
special emphasis 
to the  large $Q^2$ and $x$ region  at HERA.
In most of the instanton calculations performed so far 
the contribution of non-zero quark modes
has been neglected. It has been argued that
 this
contribution is proportional  to
the  product $m_q\rho_c$, which is indeed small for 
light $u$ and $d$ quarks.
However,  e.g. for  strange quarks, this value amounts to 
$m_s\rho_c\approx 0.25$. Evidently
 for the heavy   quarks $c$,  $b$ and $t$, the contribution of non-zero 
quark modes
in the instanton field will  dominate over the contribution 
from zero modes.

We  show that the contribution of instanton interactions
to unpolarized 
structure functions
for these modes become increasingly important at large 
 $Q^2$. 
Hence  they can provide  a QCD mechanism which leads to an  
any excess seen above calculations using parton distributions which
result from 'standard' 
perturbative QCD evolution.
For a quantitative prediction  of these 
contributions  we will use   gas 
instanton model approximation for the QCD vacuum.
In section 2 we calculate the instanton induced contribution
to the quark structure functions, using the Green functions formalism.
In section 3 we study the structure of the interaction using the 
effective Lagrangian formalism. 
 The calculation of the non-zero modes contribution 
in framework of the  liquid instanton model \ct{shur,diak} will be the 
subject of a future study.

\section{ Instanton induced contribution to the quark
structure functions}
\vspace{1mm}
\noindent

To estimate the instanton induced contribution from non-zero quark 
mode states
 to the nucleon structure functions
$F_2(x,Q^2)$, $F_L(x,Q^2)$  we  consider the  contribution to
valence quark structure function which is given by  the diagrams in Fig.1.

\begin{figure}[htb]
%\begin{figure}
%\centering
%\hspace*{5cm}\mbox{\epsfig{file=corr.eps,height=12cm,angle=0}}
\hspace*{5cm}\mbox{\epsfig{file=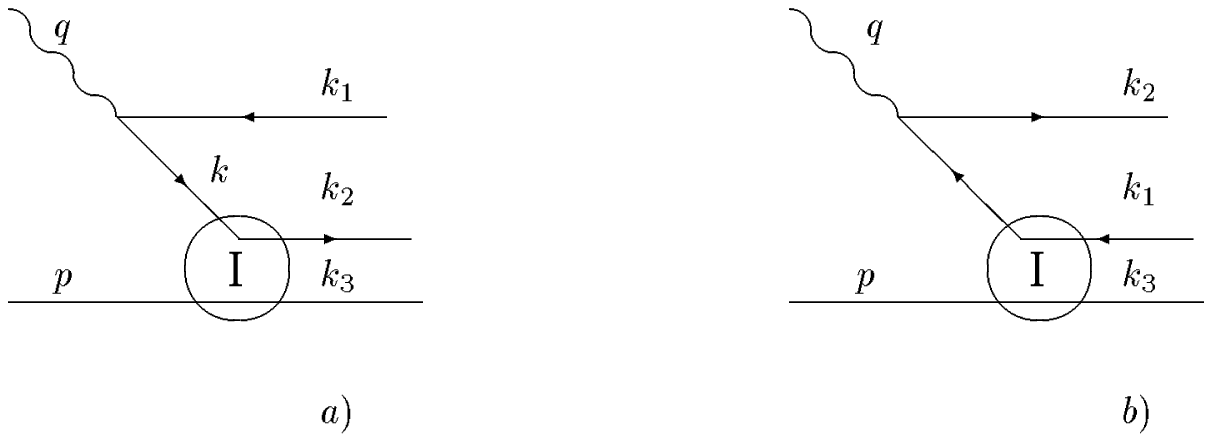,width=10cm}}
\vskip 4cm
\caption{\it   The contribution from sea quarks created by instantons
to the valence quark structure functions, a) the interaction
of the sea quark with the valence quark and b) the interaction of the
sea antiquark with the valence quark.}
\end{figure}

This contribution is determined by the deviation of  the quark Green
function in the  instanton field (i.e. the quark propagator in the 
background colour instanton field)
from a free quark Green function $S_0(x,y)$.
\be
S^{int}(x,y)=S_{I}(x,y)-S_0(x,y),
\la{int1}
\ee
with $x, y$ Euclidean space coordinates, and
where $S_I(x,y)$ is the quark Green function in an instanton field and
\be
S_0(x,y)=-\frac{\gamma\cdot(x-y)}{2\pi^2(x-y)^4}.
\ee
 The function $S_I(x,y)$ for a quark with current mass $m$ can be presented
 as the sum of the zero quark modes
and non-zero quark modes contribution (see \ct{shur})
\footnote{ We will present most formulae  in Euclidean space.
Only in the final step of the calculation the analytical continuation to
the Minkowskii space will be performed.}
\be
S_I(x,y)=S_z(x,y)+S_{nz}(x,y)=\frac{\Psi_0(x)\Psi_0^+(y)}{im}+\sum_{\lambda\neq 0}
\frac{\Psi_\lambda(x)\Psi^+_\lambda(y)}{\lambda+im},
\ee
where
\be
{\Psi_0}_{i,\alpha}(x){\Psi_0^+}_{j,\beta}(y)=\frac{1}{8}\phi(x)\phi(y)
(\hat x
\gamma_\mu\gamma_\nu\hat y\gamma_-)_{ij}\otimes(U\tau_\mu^-
\tau_\nu^+U)_{\alpha\beta},
\la{zm}
\ee
\be
\phi(x)=\frac{\rho}{\pi}\frac{1}{\sqrt{x^2}(x^2+\rho^2)^{3/2}}
\nn
\ee
and
\ba
S_{nz}(x,y)&=&\frac{1}{\sqrt{1+\rho^2/x^2}}\frac{1}{\sqrt{1+\rho^2/y^2}}(
S_0(x,y)(1+\frac{\rho^2U\tau^-\cdot x\tau^+\cdot yU^+}{x^2y^2})\nn\\
&-&D_0(x,y)\frac{\rho^2}{x^2y^2}(\frac{U\tau^-x\tau^+\cdot\gamma
\tau^-\cdot (x-y)\tau^+\cdot yU^+}{\rho^2+x^2}\gamma_++\nn\\
& &\frac{U\tau^-\cdot x\tau^+\cdot(x-y)\tau^-\cdot\gamma\tau^+\cdot yU^+}
{\rho^2+y^2}\gamma_-)) + O(m),
\la{nzm}
\ea
where $i,j$ ($\alpha,\beta$) are the spinor (colour)  indices,
$\gamma_{\pm}=(1\pm\gamma_5)/2$,  $\tau^\pm_\mu=(\vec\tau,\mp)$ is
 the colour matrix,   $U$ is  the orientation matrix of instanton in the
colour space, 
$\rho$ is the instanton size, and
\be 
D_0(x,y)=\frac{1}{4\pi^2(x-y)^2}
\la{scal}
\ee
is the propagator of a scalar quark.

The contribution of zero quark modes to  the polarized structure
function $g_1(x)$ and the unpolarized structure function 
$F_2$ has been estimated   in~\ct{kochprd} and \ct{kochhq}.
It was shown that this contribution is  rather large, specially 
at low $x$, can  
explain the violation of the Ellis-Jaffe sum rule and 
 gives a contribution to the high $Q^2$ region at HERA, on top of
the perturbative QCD contribution.

However, as discussed in section 1,
in the large $x$ and $Q^2$ region one can expect also large contributions 
from  non-zero quark modes, which are not 
very strongly localized in the instanton field and therefore their 
contribution is expected to have a different $x$ and $Q^2$ dependence 
compared to the  one from zero quark modes.

We will estimate this contribution from  non-zero quark modes
for the case of   two flavor quark 
$N_f=2$,  where one quark
will be heavy. In this case the dominated contribution 
  is related to the one shown in 
the diagrams of Fig.1, where 
the sea quark is in a non-zero mode state and the valence quark is in a
zero mode state.

The contribution to the valence quark part of 
the  structure
functions $F_2^v(x,Q^2)$ and $F_L^v(x,Q^2)$ is given by 
\ba
F_2^v(x,Q^2)&=&(-g_{\mu\nu}+6x\frac{p_{\mu}p_{\nu}}{p\cdot q})
xW_{\mu\nu}(q,p),\nn\\
F_L^v(x,Q^2)&=&4x^2\frac{p_{\mu}p_{\nu}}{p\cdot q}W_{\mu\nu}(q,p),
\la{dif}
\ea
where $p$ is the momentum of 
the valence quark, $q$ is the virtual photon momentum,
$x=Q^2/2p\cdot q$ and
\be
W_{\mu\nu}(p,q)=\frac{1}{4\pi}\prod_{i=1}^3\int \frac{d^4k_i}{(2\pi)^3}
(2\pi)^4\delta(k_i^2-m_i^2)\delta(q+p-k_1-k_2-k_3)T_\mu(p,q)T_\nu^* (p,q),
\la{tenzor}
\ee
where $T_\mu(p,q)$ is  the matrix element of the  virtual photon-valence 
quark scattering.

The instanton--induced non-zero quark mode contribution in leading order
in the mass expansion, for  sea quarks with  mass $m_q$, is
\ba
T_\mu(p,q)&=&-ie_q\int_0^{\rho_{cut}} \frac{d(\rho)d\rho}{\rho^5}
\int d^4ye^{-iqy}\bar s(k_2)\hat k_2
(\bar S_{nz}^{int}
(y,k_2)\gamma_\mu S_{nz}(k_1,y)+\nn\\
& &\bar S_{nz}
(y,k_2)\gamma_\mu  S_{nz}^{int}(k_1,y))\hat k_1 s(k_1)\bar u(k_3)
\hat k_3 S_z(k_3,p)\hat p
u(p),
\la{ampl}
\ea
where  $s(k_i)$ and $u(p)$ are Dirac spinors for sea quarks  and valence quark
respectively,
and $\rho_{cut}$ is a  parameter  which  cuts the contribution
from  large instantons.
 The density of the instantons
is
\be
d(\rho)=\frac{C_1}{2}e^{-3C_2+N_fC_3}\left(\frac{2\pi}{\alpha_s(\mu)}
\right)^6
e^{-\frac{2\pi}{\alpha_s(\mu)}}(\rho\mu)^b\prod_{q=1}^{Nf}(m_q\rho),
\la{den}
\ee
where
\be
b=b_0+\frac{\alpha_s(\mu)}{4\pi}(b_1-12b_0),
\nn
\ee

\ba
\alpha_s(\mu)=\frac{4\pi}{b_0\ln(\frac{\mu^2}{\Lambda^2})}
\left\{1-\frac{b_1}{b_0^2}\frac{\ln\ln(\frac{\mu^2}{\Lambda^2})}
{\ln\frac{\mu^2}{\Lambda^2}}\right\},
\nn
\ea
and in the $\overline{MS}$-scheme 
we have $C_1=0.466$, $C_2=1.54$ and $C_3=0.153$, where
$\mu$ is the renormalization scale, and
\be
b_0=11-\frac{2}{3}N_f{\ }{\ } b_1=102-\frac{38}{3}N_f.
\nn
\ee
For numerical estimations below the  value $\Lambda=230 MeV$ was taken.

The Fourier transform of the Green function for non-zero mode states
for quarks  is
\be
S_{nz}(y,k_2)=\int d^4xe^{ik_2x}S_{nz}(y,x)
\nn
\ee
and antiquarks is
$\bar S_{nz}(k_1,y)$ for the case $k_1^2\rightarrow 0$,
$k_2^2\rightarrow 0 $ has been calculated  in \ct{ring}
\ba
S_{nz}(y,k_2)\hat k_2&=&-\frac{iy}{\sqrt{y^2+\rho^2}}e^{ik_2y}\left[1+
\frac{\rho^2}{2y^2}\frac{U\tau^-\cdot y\tau^+\cdot k_2U^+}{k_2\cdot y}
(1-e^{-ik_2y})\right],\nn\\
\hat k_1\bar S_{nz}(k_1,y)&=&\frac{iy}{\sqrt{y^2+\rho^2}}e^{ik_1y}\left[1+
\frac{\rho^2}{2y^2}\frac{U\tau^-\cdot k_1\tau^+\cdot yU^+}{k_1\cdot y}
(1-e^{-ik_1y})\right].
\la{ftnz}
\ea

The Fourier transformed Green function for 
 zero quark modes in an instanton field equals
\be
\hat k_3 S_z(k_3,p)\hat p
=\frac{\rho^2\pi^2}{2}(\gamma_{\mu^\prime}\gamma_{\nu^\prime}\gamma_-)
\otimes(U\tau_{\mu^\prime}^-\tau_{\nu^\prime}^+U^+)F(p^2)F(k_3^2),
\la{ftz}
\ee
where
\be
F(k^2)\approx e^{-\rho|k|}\rightarrow 1{\ } at {\ }k^2\rightarrow 0.
\la{conf}
\ee

The result of the calculation  of \re{ampl}, using \re{ftnz},\re{ftz}
at $k_i^2\rightarrow 0$ is
\ba
T_\mu(p,q)&=&-e_q\pi^4\int_0^{\rho_{cut}} \frac{d(\rho)d\rho}{\rho}
\bar s(k_2)\gamma_\mu s(k_1)F_I(k,k_1,k_2)\cdot\nn\\
 & &  \frac{\bar u(k_3)(\gamma_{\mu^\prime}
\gamma_{\nu^\prime}\gamma_-)\otimes(U\tau_{\mu^\prime}^-\tau_{\nu^\prime}^+U^+)
u(p)}{m_u},
\la{ampl1}
\ea
where $k=q-k_1-k_2$ and
\ba
F_I(k,k_1,k_2)&=&I_1(k^2)+ \frac{U\tau^-\cdot k_2\tau^+\cdot kU^+}{k_2\cdot k}I_2(k,k_1,k_2)+\nn\\
& &\frac{U\tau^-\cdot k_2\tau^+\cdot k_1U^+}
{k_2\cdot k_1}I_3(k,k_1,k_2)+\frac{U\tau^-\cdot k\tau^+\cdot k_1U^+}
{k_1\cdot k}I_4(k,k_1,k_2)+k_1\leftrightarrow k_2,
\nn
\ea
with
\ba
I_1(k^2)&=&\frac{2}{\rho^2k^4}\int_0^\infty
\frac{dzz^3(z-\sqrt{z^2-\rho^2k^2})}{z^2-\rho^2k^2}J_1(z) ,\nn\\
I_2(k,k_1,k_2)&=&
\rho\left\{\frac{K_1(\sqrt{-(k+k_2)^2}\rho)}{\sqrt{-(k+k_2)^2}}-
\frac{K_1(\sqrt{-k^2}\rho)}{\sqrt{-k^2}}\right\},\nn\\
I_3(k,k_1,k_2)&=&\rho^2k_2\cdot k_1\int_0^1
d\beta\frac{K_0(\sqrt{-(k+k_1+\beta k_2)^2}\rho)-K_0(\sqrt{-(k+\beta k_2)^2}
\rho)}{2(k\cdot k_1+\beta k_1\cdot k_2)} ,
\nn\\
I_4(k,k_1,k_2)&=&\int_0^\infty dzz 
\left\{\frac{1}{-(k+k_1)^2(z^2-\rho^2(k+k_1)^2)}+\right.\nn\\
& &\left.\frac{1}{k^2(z^2-\rho^2k^2)}\right\}J_1(z).
\la{int}
\ea

To get the final result we have to  integrate over
the colour orientation $dU$ which is given by 
\ba
\int dU U_{ij}U^+_{kl}&=&\frac{1}{N_c}\delta_{jk}\delta_{li}\nn\\
\int dU U_{ij}U^+_{kl}U_{mn}U^+_{op}&=&\frac{1}{N_c^2}\delta_{jk}\delta_{li}
\delta_{no}\delta_{mp}+\frac{1}{4(N_c^2-1)}(\lambda^a)_{jk}
(\lambda^a)_{li}(\lambda^b)_{no}(\lambda^b)_{mp}.
\la{aver}
\ea

In this paper only  the colour singlet exchange between quarks
induced by instantons is taken into account. The contribution of
colour octet exchange is suppressed by a factor 
$1/N_c$ and will be studied in a 
forthcoming  paper. For the colour singlet case the final result for the
non-zero quark modes contribution to the valence 
quark structure functions $F_2$ and $F_L$,
after averaging over the initial quark polarization and colour, equals
\ba
F_2^v(x,Q^2)&=-&\frac{2^9\pi^7e_q^2x}{27m_u}
\int dPS^{3}\int_0^{\rho_{cut}} d\rho \frac{d(\rho)}
{\rho}\bar F_I^2(k,k_1,k_2)t^2\left(k_1\cdot k_2+
\frac{12x^2k_1\cdot pk_2\cdot p}{Q^2}\right),\nn\\
F_L^v(x,Q^2)&=&-\frac{2^{12}\pi^7e_q^2x^3}{27m_u}\int dPS^{3}
\int_0^{\rho_{cut}} d\rho \frac{d(\rho)}
{\rho}\bar F_I^2(k,k_1,k_2)t^2
\frac{k_1\cdot pk_2\cdot p}{Q^2},
\la{fin}
\ea
 where $dPS^{3}$ is the phase space integration,  $t^2=(p-k_3)^2$,
and 
\be
\bar F_I(k,k_1,k_2)=I_1(k^2)+ I_2(k,k_1,k_2)+I_3(k,k_1,k_2)+I_4(k,k_1,k_2)
+k_1\leftrightarrow k_2,
\nn
\ee
The integration is  performed in the center of mass frame
of the sea quarks.
In this frame the momenta of particles are (see for example \ct{shul})
\ba
q&=&(Q_0,0,0,E_q)\nn\\
t&=&(t_0,0,0,-E_q)\nn\\
p&=&E_p(1,\sin\beta,0,\cos\beta)\nn\\
k_{1,2}&=&(E_k,\mp q_k\sin\theta \cos\Phi,\mp q_k\sin\theta \sin\Phi,\mp 
q_k\cos\theta),
\la{mom}
\ea
where 

\ba
 E_k&=&\sqrt{\hat s}/2,{\ } q_k=\sqrt{\hat s-4m_q^2}/2 {\ }\nn\\
 Q_0&=&(\hat s-Q^2-t^2)/4E_k,{\ } t_0=2E_k-Q_0,{\ }\nn\\ 
E_q&=&\sqrt{(\hat s+Q^2+t^2)/4\hat s-t^2},{\ }S_G=(q+p)^2,\nn\\
E_p&=&(S_G+Q^2+t^2)/4E_k,{\ }\hat s=(t+q)^2,\nn \\
 \cos\beta&=&(-S_G-Q^2+(S_G+Q^2+t^2)Q_0/2E_k)/(2E_pE_q).\nn\\
\ea
For the phase space integration we have 
\ba
dPS^{3}=\frac{x}{128\pi^3Q^2}\int_{S_0}^{S_G}d\hat s
\int_{0}^{{t^2}_{max}}
dt^2
\int_{{t_1^2}_{min}}^{{t_1^2}_{max}}\frac{dt_1^2}{\hat s-t^2+Q^2}
\frac{d\Phi}{2\pi},
\la{ph}
\ea
where $t_1^2=(q-k_1)^2$, $S_0=(2m_q)^2$, $S_G=Q^2(1-x)/x$,
 $t_{max}^2=-(S_G-\hat s)/(1-x)$, and
\be
t_1^{max,min}=-\frac{\hat s+Q^2-t^2}{2}\left\{1\pm \sqrt{(1-
\frac{4m_q^2}{\hat s})(1+\frac{4Q^2t^2}
{(\hat s+Q^2-t^2)^2})}\right\}.
\nn
\ee

In the calculation we  take for the renormalization scale
$\mu = 1/\rho_{cut}$. For most calculations 
there is a very strong dependence on the
cutoff scale $\rho_{cut}$ of the instanton density
 on instanton induced  contributions to physical
quantities.
 From our point of view a natural
way to derive an estimate for this parameter is by considering the
confinement of the valence quarks inside the proton. 
For a zero mode valence quark in the instanton field we 
thus expect a  cutoff factor $exp(-2|p|\rho)$, where 
$|p|\approx 300 MeV$ represents the average
virtuality of a valence quark in the proton. This leads to a value
of
$\rho_{cut}\approx 1.6 GeV^{-1}$, which happens to coincide with the 
average size of the
instantons in QCD vacuum (see \ct{shur}).

\begin{figure}[htb]
%\begin{figure}
\centering
\epsfig{file=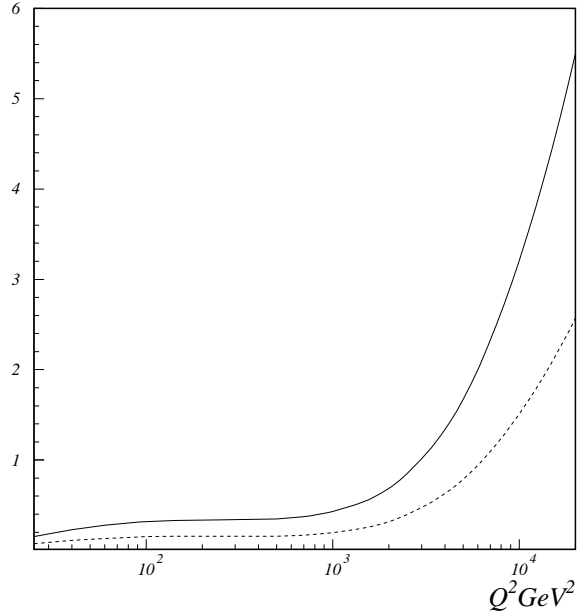,width=9cm}
\vskip 1cm
\caption{\it The instanton--induced charm quark pair contribution to 
 the  valence quark  structure functions
$F_2^v(x,Q^2)$ at $x=0.5$ (full line) and $F_L^v(x,Q^2)$ (dashed line).
at $x=0.5$, for a  charm quark mass of $m_c=1.5GeV$.}
\end{figure}

In  Fig.2  we present the result of the 
calculation of the instanton induced contribution
from the  non-zero  charm quark modes to the structure functions
$F_2^v(x,Q^2)$ and
$F_L^v(x,Q^2)$ at  $x=0.5$. This is   approximately the  $x$ region
 where the  HERA experiments have reported an excess at
high $Q^2$.
In the gas instanton model and for case $N_f=2$, as studied here, 
the strange quark contribution to the $F_2^v$ and $F_L^v$ structure 
functions from non-zero quark modes is 
expected to be about a factor  $10^{-2}$ suppressed compared to the 
one
of the charm quark, due to the smaller current quark mass.
However, in the framework of the perhaps more realistic 
liquid instanton model for the QCD vacuum \ct{shur}, \ct{diak} an
additional enhancement of strange quark production 
is expected, due to  chiral symmetry
breaking which leads to an increase of the effective quark mass.

\begin{figure}[htb]
%\begin{figure}
\centering
\epsfig{file=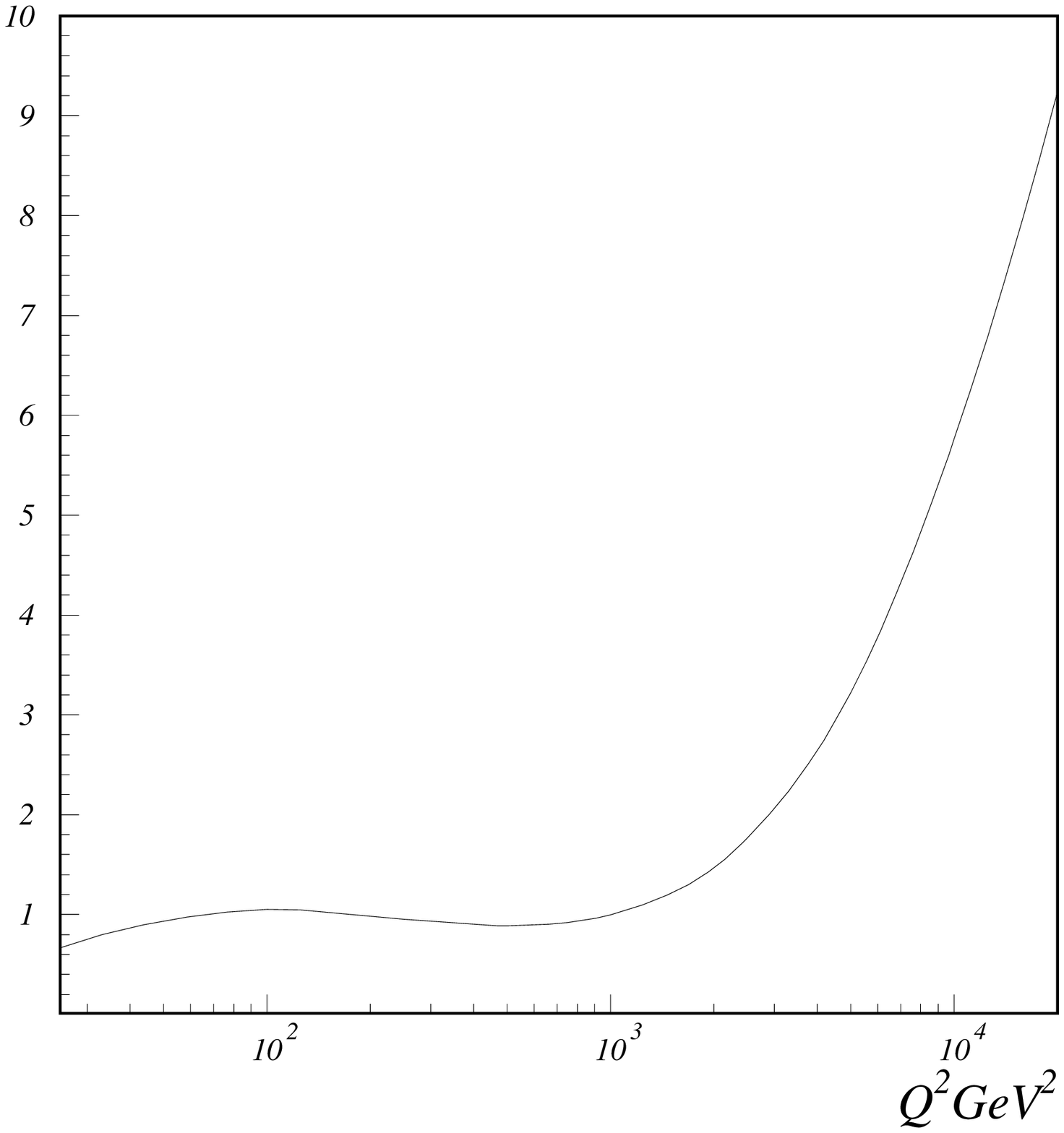,width=9cm}
\vskip 1cm
\caption{\it  The ratio of  the instanton
induced charm quark pair contribution to the valence quark  structure function
$F_2^v(x,Q^2)$, to the 
  perturbative QCD contribution,  
at $x=0.5$.}
\end{figure}

 In Fig.3 the result of the calculation of 
instanton contribution to the $F_2(x,Q^2)$ 
structure function is compared with the perturbative QCD
contribution to the same structure
function, due to gluon emission
\be
F_2^{v,pert}(x,Q^2)=e_u^2x\frac{\alpha_s(\mu)}{2\pi}
P_{qq}(x)\ln(\frac{Q^2}{\mu^2}),
\la{pert1}
\ee
where
\be
P_{qq}(x)=\frac{4}{3}\frac{1+x^2}{1-x}.
\la{pert}
\ee
The instanton contribution to the parton distributions
appears to be as large as
as the perturbative evolution correction from gluons. This suggests that  
this effect should be taken into account
in the $Q^2$ analysis of the proton $F_2^p(x,Q^2)$ structure 
function  at a large $x$  and $Q^2$.

The result  for the instanton induced non-zero quark
modes contribution   is
proportional to  the square of the mass of the quark which
 is in the non-zero mode
state.
%This factor comes from instanton density which is proportional
%to product of the quark masses. 
In our calculation we have taken only into account  terms of order
$m_c^2$,  but we did  not explicitly 
take the sea quark  mass in the Green function into account.
\footnote{ Threshold effect due to the quark masses are
taken into account  in the phase
space integration.}.
At small $Q^2\approx 20 GeV^2$  this contribution is rather small
due to phase space restriction. But for large $Q^2$ the contribution
to both $F_2^v(x,Q^2)$ and $F_L^v(x,Q^2)$ becomes very large due to 
the strong
$Q^2$ dependence of this contribution.
 This strong $Q^2$ dependence of the
instanton contribution  originates from the dimensional effective coupling 
of the instanton--induced interaction which is given by some power
 of the instanton size $\rho$. 
Therefore, similar to the famous Fermi 
effective weak interaction, the interaction through instantons should 
in principle lead 
to a violation of the unitarity bound for the DIS cross section.
However, additional perturbative and nonperturbative 
QCD corrections to  the bare instanton vertex, as discussed in 
in \ct{mul,zah},
are expected to  cancel the anomalous growth of the instanton 
contribution to DIS structure functions at very large values of $Q^2$.

 It is  interesting to note
that the non-zero quark modes also lead to a sizable contribution to
the  $F_L^v(x,Q^2)$
structure function at large $Q^2$. This results from  the  
structure of the Green function for quarks in the instanton field which
includes a part that comes from   to the  Green function of  scalar
quark (see \re{nzm}, \re{scal}).

Therefore the charm quark contribution induced by instantons to 
structure functions is large. We also expect  some contribution 
from  $b$ quarks due to the same mechanism.
Experimentally this effect could be
 verified by studying the evolution of the 
charm component of the structure function. A considerable increase
with increasing $Q^2$ is predicted.

While this study was obviously stimulated by the excess of the high
$Q^2$ events seen at HERA, it should be noted that independently 
of it's confirmation in future, this type of contribution can play an
important role in QCD analyses of very high $Q^2$ data, and should 
therefore be considered. In a future paper we plan to make more complete 
study of its implications for QCD studies.

\section{ Effective quark-quark interaction induced by
 non-zero quark modes in instanton field}
\vspace{1mm}
\noindent

In the previous section a striking $Q^2$ dependence of
the non-zero quark modes contribution to the structure functions
has been found. This  $Q^2$ dependence
results from  the structure of
quark-quark interaction vertex induced by non-zero quark modes
in the instanton field.
Let us consider an instanton induced two quark interaction 
for the case when one of the incoming quarks is off-shell, a
 situation which is directly related to the sea quark contribution
to the structure function.
One  part of the interaction is related to the
t'Hooft interaction induced by quark zero mode states.
The effective vertex can obtained by amputating of the four-point
zero modes quark Green function in instanton field
\be
{\cal L}_z=\int\frac{d\rho d(\rho)}{\rho^5}\bar
 u(k_2)\hat k_2 S_z(k_2,k_1)\hat k_1 u(k_1)
\bar d(k_3)\hat k_3 S_z(k_3,p)\hat p d(p).
\la{z1}
\ee
After substitution the Fourier transformed zero modes \re{ftz} and
performing the integration over the instanton orientation in
colour space $dU$, one  obtains the following 
effective Lagrangian for   the 
t'Hooft interaction
\be
{\cal L}_z=\int\frac{d\rho d(\rho)}{m_um_d\rho^5}
 (\frac{4\pi^2\rho^2}{3})^2
(\bar u_{R}(k_2)u_{L}(k_1)\bar d_{R}(k_3)d_{L}(p)
(1+\frac{3}{8}(1-\frac{3}{4}\sigma_{\mu\nu}^u\sigma_{\mu\nu}^d)
t_u^at_d^a+(R\longleftrightarrow L)))F(|k_1|\rho),
\la{hooft}
\ee
where $F(|k_1|\rho)\approx exp(-|k_1|\rho)$ for $\rho\rightarrow 0$.
The density of the instantons is proportional
to the product of the quark masses  \re{den} and therefore,  in spite of 
the quark masses in the denominator of Eq.\re{hooft}, the final
result is finite in the limit $m_u,m_d\rightarrow 0$.

The effective Lagrangian connected with the non-zero modes contribution
to leading order in $m_q$,  and for a small 
instanton size $\rho\rightarrow 0$ can be obtained
in a similar way
 \be
{\cal L}_{nz}=\int\frac{d\rho d(\rho)}{\rho^5}\bar s(k_2)\hat k_2
\left\{S_{nz}(k_2,k_1)-S_0(k_2,k_1)\right\}\hat k_1 s(k_1)
\bar d(k_3)\hat k_3 S_z(k_3,p)\hat p d(p).
\la{z}
\ee

 The final result is
\ba
{\cal L}_{nz}&=&\int\frac{d\rho d(\rho)}{m_d\rho^5}
 (\frac{4\pi^2\rho^2}{3})^2\left\{(1+\frac{3}{32}t^a_st_d^a)(\overline{F^n_1}(t^2)+
\overline{F^n_2}(k_1^2,t^2))\right.\nn\\
&-&\left.\frac{9}{32}\frac{t^a_st^a_dk_2^\tau k_1^\sigma{
\sigma_{\tau\sigma}}^d}{k_2\cdot k_1}
\overline{F^n_2}(k_1^2,t^2)\right\}\otimes\bar s(k_2)
\hat k_1 s(k_1)\bar d(k_3)d(p),
\la{nz}
\ea

where $t^2=(k_1-k_3)^2$ and the form factors
\ba
\overline{F^n_1}(t^2)&=&\frac{2}{\rho^2t^4}\int_0^\infty dz
\frac{z^2(z-\sqrt{z^2-\rho^2t^2})}{\sqrt{z^2-\rho^2t^2}}J_1(z),\nn\\
\overline{F^n_2}(k_1^2,t^2)&=&\int_0^\infty dzz\left\{\frac{1}{k_1^2\sqrt{z^2-k_1^2\rho^2}}-
\frac{1}{t^2\sqrt{z^2-t^2\rho^2}}\right\}J_1(z).
\la{ff}
\ea

A very interesting feature of this  interaction\re{nz} is that it is
zero for on-shell massless quarks but it is  large for the case
where one of the quarks is off-shell. This is the origin of the 
striking $Q^2$ dependence of the contribution of this interaction
to the DIS structure functions, as found in the previous section.
Another feature of the non-zero mode  interaction, as
 compared to the 
zero mode interaction \re{z} is the  helicity flip.
The  t'Hooft interaction leads to a
double spin flip of quarks and
therefore this interaction can give a large contribution to the
double spin asymmetry in the polarized DIS (see \ct{kochprd}).
The contrary is true for  non-zero quark mode
interactions, which do not have a definite
helicity and can lead only to a single spin flip; their
contribution to the double spin asymmetry is {\it zero }.
Hence these instanton induced interactions do  not 
contribute to  the spin-dependent structure function $g_1(x,Q^2)$.
The zero quark mode interactions  on the other hand 
do contribute to  $g_1(x,Q^2)$, and
this contribution is {\it negative} \ct{kochhq}. 
Thus
the total contribution of instantons to $g_1(x,Q^2)$ is negative,
contrary to the perturbative QCD expectation which yield a
 {\it positive} value for $g_1(x,Q^2)$ at large $x$ and $Q^2$.
Hence 
the instanton mechanism for DIS events excess at HERA
can be checked  by a measurement of the double spin asymmetry for these
events at high $Q^2$, 
when a  polarized proton beam at HERA will be available.
The zero quark mode component will contribute with a negative asymmetry,
while the non-zero quark mode is predicted to reduce the asymmetry,
compared to the positive asymmetry  predicted from perturbative QCD.

\section{Summary}
\vspace{1mm}
\noindent
In summary, a new contribution to the proton structure at 
high $Q^2$ and high $x$ is suggested.  
 This contribution is connected with complicated structure 
of the QCD vacuum, related to the existence of strong 
fluctuations of the vacuum gluon fields: instantons.  
It is shown  that a new type of quark-quark interaction in QCD,
which is due to non-zero quark mode states in the instanton field,
gives  a large  heavy quark contribution 
to the quark structure functions
in the large $Q^2$, $x$ region. This effect could already be visible 
at HERA.

The large  $Q^2$ dependence of this interaction comes from the
dependence of the effective instanton--induced vertex on the  quark
virtuality.
We also predict  anomalous spin properties  of high $Q^2$ events
induced by this new interaction.
 Due to the strong dependence of this interaction
on the  quark virtuality it  should show up also 
in other 
lepton-hadron and hadron-hadron  processes  with large transfer momentum. 
Note that this contribution is missing in standard QCD analyses.
Independently of a possible excess at large $Q^2$ and $x$ of the 
HERA data this contribution may have to be 
be considered when evolving to 
large scales.
Finally we note  that enhancement of the charm contribution
to the structure function at large $x$ can lead  also to an  explicit
QCD model for intrinsic charm \ct{brodsky} in the proton.
%, which 
%may be required 
%to account the   charm production in DIS at low $Q^2$
%and hadron-hadron  interactions.

\section{Acknowledgements}
\vspace{1mm}
\noindent
One of the authors (N.K.) is
thankful to   A.E.Dorokhov and S.B.Gerasimov
for many stimulated discussions.

This work was support in part by the Heisenberg-Landau program and by the
Russian Foundation for Fundamental Research (RFFR) 96-02-18096.

\end{document}